\documentstyle[12pt]{article}
\begin{document}
\tolerance=5000
\def\be{\begin{equation}}
\def\ee{\end{equation}}
\def\bea{\begin{eqnarray}}
\def\eea{\end{eqnarray}}
\def\nn{\nonumber \\}
\def\cF{{\cal F}}
\def\det{{\rm det\,}}
\def\Tr{{\rm Tr\,}}
\def\e{{\rm e}}
\def\etal{{\it et al.}}
\def\erp2{{\rm e}^{2\rho}}
\def\erm2{{\rm e}^{-2\rho}}
\def\er4{{\rm e}^{4\rho}}
\def\etal{{\it et al.}}
\def\gsim{\ ^>\llap{$_\sim$}\ }

\def\SEH{S_{\rm EH}}
\def\SGH{S_{\rm GH}}
\def\AdS5{{{\rm AdS}_5}}
\def\S4{{{\rm S}_4}}
\def\gfv{{g_{(5)}}}
\def\gfr{{g_{(4)}}}
\def\SC{{S_{\rm C}}}
\def\RH{{R_{\rm H}}}

\def\wlBox{\mbox{
\raisebox{0.1cm}{$\widetilde{\mbox{\raisebox{-0.1cm}\fbox{\ }}}$}}}
\def\htBox{\mbox{
\raisebox{0.1cm}{$\hat{\mbox{\raisebox{-0.1cm}{$\Box$}}}$}}}

\ 

\vskip -3cm

\  \hfill
\begin{minipage}{2.5cm}
December 2001 \\
\end{minipage}

\vfill

\begin{center}
{\large\bf 
Wilson loop and dS/CFT correspondence
}

\vfill

{\sc Shin'ichi NOJIRI}\footnote{nojiri@cc.nda.ac.jp} 
and {\sc Sergei D. ODINTSOV}$^{\spadesuit}$\footnote{
odintsov@ifug5.ugto.mx, odintsov@mail.tomsknet.ru} \\

\vfill

{\sl Department of Applied Physics \\
National Defence Academy,
Hashirimizu Yokosuka 239-8686, JAPAN}

\vfill

{\sl $\spadesuit$
Tomsk State Pedagogical University, Tomsk, RUSSIA and
Instituto de Fisica de la Universidad de Guanajuato, \\
Lomas del Bosque 103, Apdo. Postal E-143, 
37150 Leon,Gto., MEXICO}

\vfill

{\bf ABSTRACT}

We calculate Wilson loop (quark anti-quark potential) in dS/CFT 
correspondence. The brane-world model is considered where 
bulk is two 5d Euclidean de Sitter spaces and boundary (brane) is 4d de 
Sitter one. Starting from the Nambu-Goto action, the calculation of the 
effective tension (average energy) is presented.
Unlike to the case of supergravity calculation of Wilson loop in AdS/CFT 
set-up, there is no need to regularize the Nambu-Goto action (the volume 
of de Sitter space is finite). It turns out that at sufficiently small 
curvature of 5d background the energy (potential) is positive and linear on 
the distance between quark and anti-quark what indicates to the 
possibility of confinement.

\end{center}

\vfill


\newpage

AdS/CFT correspondence \cite{AdS} is the duality 
 between $(d+1)$-dimensional supergravity in the AdS 
background and $d$-dimensional CFT living on its boundary.
 In particulary, 
AdS$_5$/CFT$_4$ 
is described by the near horizon limit of D3-branes in Type IIB 
string theory. Hull has shown \cite{Hull} that if we consider Type 
IIB$^*$ string theory, which is given by the time-like T-duality 
from Type II string theory, D-brane is transformed into E-brane 
and its near light-cone limit gives the de Sitter space instead of 
the AdS space, which is the near horizon limit of D-brane. This 
indicates to the existence of  dS/CFT correspondence.

Moreover, the universality of holographic principle suggests
that AdS/CFT correspondence should not be something specific.
Indeed, despite some technical problems the formulation of
 dS/CFT correspondence has been recently proposed \cite{strominger}
 (for earlier attempts in this direction, see \cite{witten}).
The large number of recent works \cite{dsBrane, ds3} has been devoted to 
exploring of dS/CFT correspondence from various points of view.
There are number of problems in dS/CFT correspondence like: (still unknown)
CFT dual to de Sitter space is expected to be Euclidean and not unitary,
the boundaries where dual CFT should live are topologically disconnected, 
etc. Nevertheless, it became clear that most of properties and phenomena of
AdS/CFT correspondence have their analogs in dS/CFT correspondence.

In the very interesting paper \cite{Mal}  
there was suggested 
the way to calculate Wilson loop (quark anti-quark potential) using dual 
supergravity description in AdS/CFT set-up. In the present letter 
we try to develop the analagous formulation to find the
 potential between quark and anti-quark in the 
framework of dS/CFT correspondence. Despite the fact that dual CFT is 
still unknown the calculation of Nambu-Goto static potential leads 
to acceptable description of Wilson loop. 

Let us  consider 5 dimensional de Sitter (dS) space as 
a bulk space
\be
\label{dS1}
ds_{\rm dS}^2 = g_{\mu\nu} dx^\mu dx^\nu 
= - dt^2 + \e^{{2t \over l}}\sum_{i=1}^4 
\left(dx^i\right)^2\ ,
\ee
We assume there is a space-like brane at constant 
$t=t_0\rightarrow +\infty$. 
In Type IIB$^*$ string theory \cite{Hull}, the length parameter 
$l$ could be given by the corresponding CFT. For AdS/CFT where dual theory 
 is ${\cal N}=4$ $SU(N)$ super Yang-Mills theory one has
\be
\label{dS1b}
l=\left(4\pi gN\right)^{1 \over 4}\ .
\ee
Here $g$ is the coupling constant of the super Yang-Mills theory.
This relation may be used as a probe in dS/CFT, just to see qualitative 
difference with AdS/CFT.

 One starts from the static potential between 
``quark'' and ``anti-quark''. Following to \cite{Mal}, 
we evaluate the  Nambu-Goto action on dS background
\be
\label{dS2}
S_{\rm NG}={1 \over 2\pi}\int d\tau d\sigma \sqrt{\det\left(\pm g_{\mu\nu}
\partial_i x^\mu \partial_j x^\nu\right)}\ .
\ee
This action could be regarded to describe the trajectory of the 
Wilson line connectiong the source(s), which can be regarded as 
(anti-)quark. 
If the trajectory is space(time)-like, we choose $+$ ($-$) sign 
in $\pm$ in (\ref{dS2}). One considers the configuration 
$x^1\equiv x={R \over \pi}\sigma$, $x^2={T \over \pi}\tau$, 
$x^3=x^4=0$ and $t=x^0(\sigma)$. Here we assume $\sigma,\tau\in 
[0,\pi]$ but $T$ is sufficiently large. 
We now define a matrix variable $M_{ij}$ 
($i,j=\tau,\sigma$) by
\be
\label{dS3}
M_{ij}=g_{\mu\nu}\partial_i x^\mu \partial_j x^\nu\ .
\ee
Then the equations of motion are given by the variation 
of the action (\ref{dS2}) with respect to $x^\mu$:
\be
\label{dS4}
\partial_i\left(\sqrt{\pm \det M}\left(M^{-1}\right)^{ij}
\partial_j x^\mu g_{\mu\nu}\right) 
+ {1 \over 2} \sqrt{\pm \det M}\left(M^{-1}\right)^{ij}
\partial_j x^\nu \partial_j x^\rho {\partial g_{\rho\nu} 
\over \partial x^\mu}\ .
\ee
In the metric  (\ref{dS1}), one has 
\be
\label{dS5}
M_{\tau\tau} = \e^{2x^0(\sigma) \over l}\left({T \over \pi}\right)
\ ,\quad M_{\sigma\sigma} = - \left(\partial_\sigma x^0
(\sigma)\right)
+ \e^{2x^0(\sigma) \over l}\left({R \over \pi}\right)\ ,\quad 
M_{\sigma\tau}=M_{\tau\sigma}=0\ .
\ee
Then the $\mu=0$ component in (\ref{dS4}) gives 
\bea
\label{dS6}
0&=&-\partial_\sigma \left\{ {\e^{x^0 \over l}\left({T \over \pi}
\right) \partial_\sigma x^0 
\over \sqrt{\pm \left(- \left(\partial_\sigma x^0
(\sigma)\right)+ \e^{2x^0(\sigma) \over l}
\left({R \over \pi}\right)^2\right)}}\right\} \nn
&& + {\e^{3x^0 \over l} \over l}\left({T \over \pi}\right)
\sqrt{\pm \left(- \left(\partial_\sigma x^0
(\sigma)\right)+ \e^{2x^0(\sigma) \over l}
\left({R \over \pi}\right)^2\right)} \nn
&& \times \left[{\left({R \over \pi}\right)^2 \over 
- \left(\partial_\sigma x^0
(\sigma)\right)+ \e^{2x^0(\sigma) \over l}
\left({R \over \pi}\right)^2} + \e^{-{2x^0 \over l}}\right]\ ,
\eea
and $\mu=1$ component 
\be
\label{dS7}
0=\partial_\sigma \left\{ {\e^{3x^0 \over l}\left({TR \over \pi^2}
\right) \over \sqrt{\pm \left(- \left(\partial_\sigma x^0
(\sigma)\right)+ \e^{2x^0(\sigma) \over l}
\left({R \over \pi}\right)^2\right)}}\right\}\ .
\ee
One can integrate (\ref{dS7}) using 
\be
\label{dS8}
 {\e^{3x^0 \over l}\left({TR \over \pi^2}
\right) \over \sqrt{\pm \left(- \left(\partial_\sigma x^0
(\sigma)\right)+ \e^{2x^0(\sigma) \over l}
\left({R \over \pi}\right)^2\right)}}=E\geq 0\ .
\ee
Then we have
\be
\label{dS9}
\left(\partial_\sigma x^0\right)^2 
=\e^{2x^0 \over l}\left({R \over \pi}\right) 
\mp \e^{6x^0 \over l}{TR \over \pi^2 E^2}\ .
\ee
In (\ref{dS9}), the minus (plus) sign corresponds to 
the case that the trajectory is time(space)-like. 
In the time-like case, the r.h.s. of (\ref{dS9}) is always 
positive and non-vanishing. Therefore $\partial_\sigma x^0$ 
does not vanish, which means there is no trajectory 
connecting two separated sources (which could be regarded as 
quark and anti-quark) on the brane. This is, of course, 
consistent since the trajectory must become space-like 
somewhere in order that the trajectory connects the two 
separated sources on the brane.
On the other hand, the space-like case (the minus sign in 
(\ref{dS9})) is also inconsistent. Indeed, brane exists at 
$x^0=t=t_0\rightarrow +\infty$, the r.h.s. in 
(\ref{dS9}) becomes negative near the brane but the  
l.h.s. is always positive. Therefore there is no space-like 
trajectory connecting the brane. 

Hence, it seems there is no consistent solution for the 
Wilson line. Having in mind that dual CFT should be Euclidean,
 we consider 
the brane in the bulk 
space with Euclidean signature. The dS space becomes 
the sphere after the Wick-rotation. Then we consider the brane 
of 4 dimensional sphere S$_4$ in the bulk 5 dimensional 
sphere S$_5$. Such a model has been recently proposed in \cite{dsBrane}, 
where the quantum creation 
of four dimensional de Sitter brane universe in five dimensional 
de Sitter spacetime as in  scenario of so-called Brane New 
World \cite{NOZ,HHR0} is investigated. 
The metric of  5 dimensional Euclidean de Sitter space 
that is 5d sphere is given by
\be
\label{dSi}
ds^2_{{\rm S}_5}=dy^2 + l^2 \sin^2 {y \over l}d\Omega^2_4\ .
\ee
Here $d\Omega^2_4$ describes the metric of ${\rm S}_4$ 
with unit radius. The coordinate $y$ is defined between 
$0\leq y \leq l\pi$. One also assumes the brane lies at $y=y_0$ 
and the bulk space is given by gluing two regions 
given by $0\leq y < y_0$. 

We start with the action $S$ which is the sum of 
the Einstein-Hilbert action $\SEH$ with positive 
cosmological constant, the Gibbons-Hawking 
surface term $\SGH$ and the surface counter term $S_1$.\footnote{
The coefficient of $S_1$ cannot be determined from the condition of 
finiteness of the action as in AdS/CFT. However, using the 
renormailzation group method developed in \cite{BVV}  this 
coefficient can be determined uniquely, see also  third paper 
in \cite{ds3}. } 
\bea
\label{Stotal}
S&=&\SEH + \SGH + 2 S_1 \ ,\nn 
\SEH&=&{1 \over 16\pi G}\int d^5 x \sqrt{\gfv}\left(R_{(5)} 
 - {12 \over l^2}\right)\ , \nn
\SGH&=&{1 \over 8\pi G}\int d^4 x \sqrt{\gfr}\nabla_\mu n^\mu 
\ ,\quad S_1= -{3 \over 8\pi Gl}\int d^4 x \sqrt{\gfr} \ .
\eea 
Here the quantities in the  5 dimensional bulk spacetime are 
specified by the suffices $_{(5)}$ and those in the boundary 4 
dimensional spacetime  by $_{(4)}$. 
The factor $2$ in front of $S_1$ in (\ref{Stotal}) is coming from 
that we have two bulk regions which 
are connected with each other by the brane. 
In (\ref{Stotal}), $n^\mu$ is 
the unit vector normal to the boundary. In (\ref{Stotal}), 
one chooses the 4 dimensional boundary metric as 
\be
\label{tildeg}
\gfr_{\mu\nu}=\e^{2A}\tilde g_{\mu\nu}
\ee 
and we specify the 
quantities with $\tilde g_{\mu\nu}$ by using $\tilde{\ }$. 

The metric of ${\rm S}_4$ with the unit radius is given by
\be
\label{S4metric1}
d\Omega^2_4= d \chi^2 + \sin^2 \chi d\Omega^2_3\ .
\ee
Here $d\Omega^2_3$ is described by the metric of 3 dimensional 
unit sphere. If one changes the coordinate $\chi$ to 
$\sigma$ by $\sin\chi = \pm {1 \over \cosh \sigma}$, 
one obtains\footnote{
If we Wick-rotate the metric by $\sigma\rightarrow it$, we 
obtain the metric of de Sitter space:
\[
d\Omega^2_4\rightarrow ds_{\rm dS}^2
= {1 \over \cos^2 t}\left(-dt^2 + d\Omega^2_3\right)\ .
\]
}
\be
\label{S4metric2}
d\Omega^2_4= {1 \over \cosh^2 \sigma}\left(d \sigma^2 
+ d\Omega^2_3\right)\ .
\ee
Then one assumes 
the metric of 5 dimensional space time as follows:
\be
\label{metric1}
ds^2=dy^2 + \e^{2A(y,\sigma)}\tilde g_{\mu\nu}dx^\mu dx^\nu\ ,
\quad \tilde g_{\mu\nu}dx^\mu dx^\nu\equiv l^2\left(d \sigma^2 
+ d\Omega^2_3\right)
\ee
and one identifies $A$ and $\tilde g$ in (\ref{metric1}) with those in 
(\ref{tildeg}). Then the actions in (\ref{Stotal}) 
have the following forms:
\bea
\label{actions2}
\SEH&=& {l^4 V_3 \over 16\pi G}\int dy d\sigma \left\{\left( -8 
\partial_y^2 A - 20 (\partial_y A)^2\right)\e^{4A} \right. \nn
&& \qquad \left. +\left(-6\partial_\sigma^2 A 
 - 6 (\partial_\sigma A)^2 
+ 6 \right)\e^{2A} - {12 \over l^2} \e^{4A}\right\} \nn
\SGH&=& {l^4 V_3 \over 2\pi G}\int d\sigma \e^{4A} 
\partial_y A \ ,\quad 
S_1= - {3l^3 V_3 \over 8\pi G}\int d\sigma \e^{4A} \ .
\eea
Here $V_3$ is the volume or area of the unit 3 sphere. 

In the bulk, one obtains the following equation of motion 
from $\SEH$ by the variation over $A$:
\be
\label{eq1}
0= \left(-24 \partial_y^2 A - 48 (\partial_y A)^2 
 - {48 \over l^2}
\right)\e^{4A} + {1 \over l^2}\left(-12 \partial_\sigma^2 A 
- 12 (\partial_\sigma A)^2 + 12\right)\e^{2A}\ ,
\ee
which corresponds to the one of the Einstein equations. 
Then one finds solution, $A_S$, which corresponds to 
the metric dS$_5$ (\ref{dSi}) with (\ref{S4metric2}). 
\be
\label{blksl}
A=A_S=\ln\sin{y \over l} - \ln \cosh\sigma\ .
\ee
On the brane at the boundary, one gets the following equation:
\be
\label{eq2}
0={48 l^4 \over 16\pi G}\left(\partial_y A - {1 \over l}
\right)\e^{4A}\ .
\ee
We should note that the contributions from $\SEH$ and $\SGH$ are 
twice from the naive values since we have two bulk regions which 
are connected with each other by the brane. 
Substituting the bulk solution $A=A_S$ (\ref{blksl}) into 
(\ref{eq2}) and defining the radius $R$ of the brane by
$R\equiv l\sin{y_0 \over l}$, one obtains
\be
\label{slbr2}
0={1 \over \pi G}\left({1 \over R}\sqrt{1-{R^2 \over l^2} }
 - {1 \over l}\right)R^4 \ .
\ee
First we should note $0\leq R\leq l$ by definition. 
We find Eq.(\ref{slbr2}) has a solution:
\be
\label{Csol}
R^2=R_0^2\equiv 
{l^2 \over 2}\ \mbox{or}\ {y_0 \over l}={\pi \over 4}, 
{3\pi \over 4}\ .
\ee
In Eq.(\ref{slbr2}), the first term 
${R^3 \over \pi G}\sqrt{1 - {R^2 \over l^2}}$ 
corresponds to the gravity, which makes the radius $R$ larger. 
On the other hand, the second term 
$-{R^4 \over \pi Gl}$ corresponds to the tension, which makes 
$R$ smaller. When $R<R_0$,  gravity becomes larger than the 
tension and when $R>R_0$, vice versa. Then both of the solutions in 
(\ref{Csol}) are stable. Although it is not clear from 
(\ref{slbr2}), $R=l$ (${y \over l}={\pi \over 2}$) corresponds 
to the local maximum. 
Hence, the possibility of creation of inflationary brane in 
de Sitter bulk 
is possible already on classical level, even in situation when brane 
tension is fixed by holographic RG. That is qualitatively different from 
the case of AdS bulk where only quantum effects led to creation of
inflationary 
brane \cite{NOZ,HHR0} (when brane tension was not free parameter).

In order to consider the potential between quark and anti-quark, 
we embedd the bulk S$_5$ in 6-dimensional flat Euclidean space, 
whose orthogonal coordinates are denoted by $\{x^i\}$ ($i=1,2,
\cdots 6$) as follows
\bea
\label{S5i}
&& x^1=l \cos\theta_1\ ,\quad x^2 = l\sin\theta_1\cos\theta_2\ ,
\quad x^3=l\sin\theta_1\sin\theta_2\cos\theta_3\ ,\nn
&& x^4=l\sin\theta_1\sin\theta_2\sin\theta_3\cos\theta_4\ ,\quad 
x^5=l\sin\theta_1\sin\theta_2\sin\theta_3\sin\theta_4\cos\phi
\ ,\nn
&& x^6=l\sin\theta_1\sin\theta_2\sin\theta_3\sin\theta_4\sin\phi\ .
\eea
Here the polar corrdinates $\theta_i$ ($i=1,\cdots, 4$) and 
$\phi$ take their values as $\theta_i\in [0,\pi]$ and 
$\phi\in [0,2\pi)$. 
We also embedd the S$_4$ brane, whose radius is given by 
$R_0={l \over \sqrt{2}}$  (\ref{Csol}), by putting 
$\theta_1={\pi \over 4}$ (\ref{S5i}):
\bea
\label{S5ii}
&& x^1={l \over \sqrt{2}}\ ,\quad x^2 = 
{l \over \sqrt{2}}\cos\theta_2\ ,
\quad x^3={l \over \sqrt{2}}\sin\theta_2\cos\theta_3\ ,\nn
&& x^4={l \over \sqrt{2}}\sin\theta_2\sin\theta_3\cos\theta_4\ ,\quad 
x^5={l \over \sqrt{2}}\sin\theta_2\sin\theta_3\sin\theta_4\cos\phi
\ ,\nn
&& x^6={l \over \sqrt{2}}\sin\theta_2\sin\theta_3\sin\theta_4\sin\phi\ .
\eea
We also put the world line of quark and anti-quark 
at $\theta_2=\theta_3={\pi \over 2}$, $\theta_4=\theta_0$ 
($\phi\in [0,2\pi)$) on the S$_4$:
\bea
\label{S5iii}
&& x^1={l \over \sqrt{2}}\ ,\quad x^2= x^3= 0\ ,\quad 
x^4={l \over \sqrt{2}}\cos\theta_0\ ,\nn 
&& x^5={l \over \sqrt{2}}\sin\theta_0\cos\phi
\ ,\quad x^6={l \over \sqrt{2}}\sin\theta_0\sin\phi\ .
\eea
One can regard that the quark and anti-quark are created at 
$\phi=0$ and annihilated at $\phi=\pi$. 
As it is sufficient if we consider the upper half sphere of 
S$_4$, we can restrict $\theta_0$ as
\be
\label{theta0}
0\leq\theta_0\leq {\pi \over 2}
\ee
The minimum area connecting the world lines of quark and 
anti-quark on the S$_4$ brane is then given by
\be
\label{S5iv}
A_1=\pi l^2 \left(1 - \cos \theta_0\right)\ .
\ee
Eq.(\ref{S5iv}) can be understood by considering S$_2$ with the 
radius $R_0={l \over \sqrt{2}}$ on the S$_4$ brane and putting the 
world line of quark and anti-quark on the S$_2$. The 
surface of the minimum area is on the S$_2$. Then $A_1$ in 
(\ref{S5iv}) corresponds to the product of the period $T$ of the 
(Euclidean) time and the distance $L$ between quark and anti-quark 
on the flat brane $TL=A_1$. Especially if we regard $L=T$, we have 
\be
\label{S5ivb}
L^2 = \pi l^2 \left(1 - \cos \theta_0\right)\ ,
\ee
or
\be
\label{S5ivc}
\cos\theta_0= 1 - {L^2 \over \pi l^2}\ .
\ee
Eqs.(\ref{theta0}) and (\ref{S5ivb}) tell that the value of $L$ 
is restricted as
\be
\label{resL}
L^2\leq \pi l^2\ .
\ee
The Nambu-Goto action (\ref{dS2}) tells 
that the action is given by the minimum area whose boundary is 
the world line of quark and anti-quark on the bulk S$_5$. 
We should note that the world line of quark and anti-quark is 
also on the S$_4$ with the maximal radius $l$:
\bea
\label{S5v}
&& x^1=-l \sin\eta_0 \sin\hat\theta_2\sin\hat\theta_3
\cos\hat\theta_4\ ,\quad 
x^2 = l\cos\hat\theta_2\ ,
\quad x^3=l\sin\hat\theta_2\cos\hat\theta_3\ ,\nn
&& x^4=l\cos\eta_0 \sin\hat\theta_2\sin\hat\theta_3
\cos\hat\theta_4\ ,\quad 
x^5=l\sin\hat\theta_2\sin\hat\theta_3\sin\hat\theta_4\cos\phi
\ ,\nn
&& x^6=l\sin\hat\theta_1\sin\hat\theta_2\sin\hat\theta_3
\sin\hat\theta_4\sin\phi\ .
\eea
Here the angle $\eta_0$ is defined by
\be
\label{S5vi}
\tan\eta_0 = - {1 \over \cos \theta_0}\ ,
\ee
and the world line of quark and anti-quark in (\ref{S5iii}) 
is given by putting $\hat\theta_2=\hat\theta_3={\pi \over 2}$ and 
\be
\label{S5vii}
\cos\hat\theta_4=\sqrt{1 + \cos^2\theta_0 \over 2}\ .
\ee
The S$_4$ in (\ref{S5v}) is given by the rotation:
\bea
\label{S5viib}
x^i&=&\sum_{j=1}^6 M_{ij}{x^j}' \nn
M_{ij}&=&\left(\begin{array}{cccccc}
\cos\eta_0 & 0 & 0 & -\sin\eta_0 & 0 & 0 \\
0 & 1 & 0 & 0 & 0 & 0 \\
0 & 0 & 1 & 0 & 0 & 0 \\
\sin\eta_0 & 0 & 0 & \cos\eta_0 & 0 & 0 \\
0 & 0 & 0 & 0 & 1 & 0 \\
0 & 0 & 0 & 0 & 0 & 1 \\
\end{array}\right)\ ,
\eea
from an S$_4$ with the maximal radius $l$, which is given 
by putting $\theta_1={\pi \over 2}$ in (\ref{S5i})
\bea
\label{S5viii}
&& {x^1}'= 0\ ,\quad {x^2}' = l\cos\hat\theta_2\ ,
\quad {x^3}'=l\sin\hat\theta_2\cos\hat\theta_3\ ,\nn
&& {x^4}'=l\sin\hat\theta_2\sin\hat\theta_3\cos\theta_4\ ,\quad 
{x^5}'=l\sin\hat\theta_2\sin\hat\theta_3\sin\hat\theta_4\cos\phi
\ ,\nn
&& {x^6}'=l\sin\hat\theta_2\sin\hat\theta_3\sin\hat\theta_4\sin\phi\ .
\eea
Then the surface with the minimum area on the bulk S$_5$ 
surrounded by the world line of quark and anti-quark exists 
in S$_2$ with the maximal radius $l$ embedded in S$_4$ in 
(\ref{S5v}). The area is given by
\bea
\label{S5ix}
A_2 &=& 2\pi l^2 \left( 1 - \cos\hat\theta_4 \right) \nn
&=& 2\pi l^2 \left( 1 - \sqrt{1 + \cos^2\theta_0 \over 2}
\right)\ .
\eea
Then the Nambu-Goto action (\ref{dS2}) is given by
\be
\label{dS2b}
S_{\rm NG}={A_2 \over 2\pi}\ .
\ee
In case of AdS/CFT correspondence, the Nambu-Goto action 
contains the divergence coming from the infinite volume of 
the AdS space. Since S$_5$, which is given by Wick-rotating 
de Sitter space in the Euclidean signature, is compact space 
and has a finite volume, there does not appear such a kind of 
divergence. 

Eqs.(\ref{S5ix}) and (\ref{dS2b}) tell that the effective 
tension $\sigma$, which is the energy in the unit length, 
should be given by the ratio of $A_1$  
(\ref{S5iv}) and $A_2$ (\ref{S5ix}) :
\be
\label{S5x}
\sigma = {A_2 \over A_1} = {1 - \sqrt{1 + \cos^2\theta_0 \over 2}
\over 1 - \cos\theta_0}\ .
\ee
Or if we use the distance $L$ between the quark and anti-quark 
in (\ref{S5ivc}), one gets
\be
\label{S5xb}
\sigma  = {\pi l^2 \left\{1 - \sqrt{1 + 
\left(1 - {L^2 \over \pi l^2}\right)^2 \over 2}
\right\}\over L^2}\ .
\ee
Then the average energy $E$ in the process of the creation and 
annihilation of quark and anti-quark is given by 
\be
\label{S5xi}
E=L\sigma 
= {\pi l^2 \left\{1 - \sqrt{1 + 
\left(1 - {L^2 \over \pi l^2}\right)^2 \over 2}
\right\}\over L}\ ,
\ee
which could be regarded as the potential between quark and 
anti-quark.
Note once more that potential is always positive 
(unlike to the case of supergravity description in AdS/CFT),
as a result of the fact that no regularization is necessary.

 For small $L\ll l$, the potential behaves as 
\be
\label{S5xii}
E\sim {L \over 2}\ ,
\ee
which is linear. It indicates that quark and anti-quark 
could be confined only if the curvature of S$_5$ is sufficiently 
small. On the other hand, even near the equator 
($\theta_0={\pi \over 2}$ or $L^2 = \pi l^2$), the 
effective tension has the value of
\be
\label{S5xiii}
\left.\sigma\right|_{L^2=\pi l^2}=1 - {1 \over \sqrt{2}}
\ee
and does not vanish. Let us start from the small loop of the 
world line of quark and anti-quark near the north pole 
and strech the loop until the loop arrives at the equator. 
Then even on the equator, the loop suffers the non-vaishing tension 
which attracts the loop to the north pole. On the other hand, 
if one starts from the south pole, the tension attracts the 
loop to the south pole. Therefore this indicates that there 
is a kind of the first order phase transition.
 The above behavior is rather different from the supergravity description in 
the AdS/CFT correspondence \cite{Mal}, where the potential is negative and
its absolute value is
proportional to the inverse of the distance between quark and 
anti-quark.  Hence, in AdS/CFT there is an indication to the screening
behaviour. 

In order to understand the interplay between supergravity description 
of Wilson loop in AdS/CFT and dS/CFT (and also for better understanding 
of dS/CFT itself) it would be extremely interesting to repeat this 
calculation on brane-world background interpolating between AdS, dS and 
Minkowskii spaces (as solution of  ref.\cite{cvetic}).

\end{document}